\documentclass[sigconf]{acmart}

\usepackage{caption}
\usepackage{graphicx}
\usepackage{xspace}
\usepackage{cleveref}
\usepackage{lipsum}
\usepackage[normalem]{ulem}
\usepackage{enumitem}
\usepackage{geometry}
\usepackage{amsmath}
\setlist[itemize]{noitemsep, topsep=0pt}
\usepackage[utf8]{inputenc}
\usepackage[T1]{fontenc}
\usepackage{adjustbox}
\usepackage{subfigure}
\usepackage{multirow}
\usepackage[normalem]{ulem}

\usepackage{booktabs}
\usepackage{tabularx}
\usepackage{makecell}
\usepackage{array}

\usepackage{soul}
\usepackage{hyperref}
\hypersetup{
    colorlinks=true,
    linkcolor=blue,
    filecolor=magenta,
    urlcolor=cyan,
}

\usepackage{siunitx}
\usepackage{tikz}

\makeatletter
\DeclareRobustCommand\onedot{\futurelet\@let@token\@onedot}
\def\@onedot{\ifx\@let@token.\else.\null\fi\xspace}

\makeatother

\AtBeginDocument{%
  \providecommand\BibTeX{{%
    \normalfont B\kern-0.5em{\scshape i\kern-0.25em b}\kern-0.8em\TeX}}}

\usepackage{pifont}

\definecolor{goodgreen}{RGB}{0,140,70}
\definecolor{warnorange}{RGB}{230,130,0}
\definecolor{badred}{RGB}{200,40,40}

\newcommand{\cmark}{\raisebox{-0.2ex}{\textcolor{goodgreen}{\Large\ding{51}}}}
\newcommand{\pmark}{\raisebox{-0.1ex}{\textcolor{warnorange}{\Large\(\blacktriangle\)}}}
\newcommand{\xmark}{\raisebox{-0.15ex}{\textcolor{badred}{\Large\ding{55}}}}

\newcommand{\system}[1]{\textit{UPG}}

\begin{document}
\title[]{Unified Pitch Graphs for Diagnosing Pitching Strategy}


\author{Kichang Lee}
\email{kichang.lee@yonsei.ac.kr}
\affiliation{%
  \institution{Yonsei University}
  \country{}
}
\author{JeongGil Ko}
\email{jeonggil.ko@yonsei.ac.kr}
\affiliation{%
  \institution{Yonsei University}
  \country{}
}


\begin{abstract}
Pitching strategy in baseball is expressed through both physical execution and the ordered context in which pitches are used, yet common representations collapse pitches into discrete types or aggregate statistics. We present Unified Pitch Graphs (UPG), a hierarchical graph representation for retrospective analysis of sequential spatiotemporal events. \system{} preserves each pitch as an exact event with reconstructed three-dimensional trajectory and context, connects consecutive pitches through directed sequence edges, and organizes the same events across semantic and temporal resolutions. A support-adaptive mechanism backs off from fine, long sequences when repeated evidence is insufficient, while retaining exact event lineage. We evaluate \system{} on 3.94 million MLB Statcast pitches from 2021–2026. Nominally identical pitch sequences exhibit distinct physical executions, and ordered structure becomes increasingly evident in longer context-conditioned paths. Support-adaptive backoff increases held-out path coverage from 18.9\% to 94.9\% while improving execution reconstruction from $R^{2}=0.495$ to 0.685. \system{} also reliably localizes controlled execution changes that discrete pitch-mix and sequence representations cannot detect. These results demonstrate that \system{} provides a traceable, multi-scale representation for identifying recurring strategy patterns without conflating retrospective associations with causal or future-performance claims.

\end{abstract}
\settopmatter{printfolios=true} 
\settopmatter{printacmref=false} 
\renewcommand\footnotetextcopyrightpermission[1]{} 
\maketitle
\section{Introduction}
\label{sec:intro}

Baseball is one of the most data-rich domains in modern sports analytics, with a long tradition of quantitative analysis and increasingly detailed tracking of individual plays~\cite{mizels2022current}. 
Modern pitch-tracking systems record millions of pitches with release conditions, velocity, movement, three-dimensional flight, plate location, game context, and batter response. 
Such data provide an opportunity to study not only player outcomes, but also the sequential physical patterns associated with those outcomes

Pitching is particularly well suited to this type of analysis since each pitch is both a physical execution and an action within an ordered interaction. 
The strategic meaning of a pitch depends not only on its nominal type, but also on how it is executed, what preceded it, and the context in which it is thrown. 
For example, two fastballs with the same pitch-type label may differ substantially in velocity, movement, release path, and trajectory, while the same fastball--breaking-ball sequence may have different effects depending on count, matchup, location, and the physical relationship between the two pitches~\cite{long2017pitchtunnels,prasad2021graphsequencing}. 
Thus, pitching strategy is naturally relational: individual pitch events are connected through observed temporal order and embedded in multiple contextual and temporal scopes.

Graph representations provide a natural way to model such sequential structure, and prior work has used directed graphs to capture pitch-order dependencies beyond independent pitch selection~\cite{prasad2021graphsequencing}. 
However, graph construction introduces an important resolution trade-off. 
Coarse representations based on pitch types or broad transition categories are compact and well supported, yet they collapse physically distinct executions. 
Conversely, defining states jointly by trajectory, location, context, and longer pitch histories quickly fragments the representation into sparsely repeated patterns, reflecting the general trade-off between longer sequential contexts and reliable statistical support~\cite{rissanen1983universal,cunial2019framework}. 
Aggregate representations may also obscure the exact games, plate appearances, and physical pitches that support a discovered pattern. 
The central challenge is therefore not simply to construct a pitch graph, but to preserve physical fidelity and event lineage while maintaining sufficient support for recurring multi-scale patterns.

To address these limitations, we propose \system{} (\textbf{U}nified \textbf{P}itch \textbf{G}raph), a hierarchical graph framework for large-scale analysis of sequential spatio-temporal pitching events.
First, \system{} preserves each pitch as an exact event with its continuous physical execution, rather than replacing it with a coarse pitch-type state.
Second, it organizes these events across pitch type, trajectory, location, and temporal scales while retaining links to the original pitches and plate appearances.
Third, to avoid the sparsity caused by long or highly detailed sequences, \system{} adaptively backs off to shorter or coarser paths when repeated support is insufficient.
Together, these components allow recurring pitching patterns to be analyzed at a supported resolution without discarding the physical events from which they were constructed.

We evaluate \system{} on 3.94 million MLB Statcast pitches from 2021 through a partial 2026 season~\cite{mlbBaseballSavant}. 
Our experiments show that nominally identical discrete sequences can contain distinct physical executions, and that meaningful ordered structure becomes more evident in longer context-conditioned paths. 
Support-adaptive backoff increases held-out path coverage from 18.9\% to 94.9\% while improving execution reconstruction from $R^2=0.495$ to $0.685$. 
We further show that the resulting hierarchy can localize execution changes that pitch-mix and discrete sequence representations cannot detect, while preserving direct lineage from aggregate findings to their supporting games, plate appearances, and individual pitches. 
We note that these results position \system{} as a graph-based diagnostic representation for data-rich sequential event analysis rather than as a causal or future-performance prediction model.

Our contributions are as follows:
\begin{itemize}[leftmargin=*]
    \item We formulate pitching-strategy diagnosis as a large-scale sequential graph problem that jointly considers physical execution, observed pitch order, game context, and temporal scope.

    \item We propose \system{}, a hierarchical attributed graph that preserves exact pitch events and sequence relations while organizing them through semantic/temporal resolutions with full event lineage.

    \item We develop a support-adaptive variable-order representation that balances fine-grained physical fidelity against the sparsity of long and highly specific strategy paths.

    \item Using large-scale MLB tracking data, we demonstrate that \system{} reveals execution and ordered structure lost by discrete representations and supports traceable multi-scale diagnosis and retrospective change localization.
\end{itemize}
\begin{table*}[t]
\centering
\scriptsize
\setlength{\tabcolsep}{2.7pt}
\renewcommand{\arraystretch}{1.08}

\caption{
Comparison of representative pitching analytics.
\cmark: explicitly modeled,
\pmark: partially supported,
\xmark: not central.
}
\label{tab:related_work_gap}

\begin{tabularx}{\textwidth}{
@{}
>{\raggedright\arraybackslash
  \hsize=1.20\hsize
  \linewidth=\hsize}X
>{\raggedright\arraybackslash
  \hsize=0.80\hsize
  \linewidth=\hsize}X
*{7}{>{\centering\arraybackslash}p{1.12cm}}
@{}
}
\toprule

\textbf{Study / line of work} &
\textbf{Primary focus} &
\makecell{\textbf{Sequence}\\\textbf{trans.}} &
\makecell{\textbf{Trajectory}} &
\makecell{\textbf{Context}} &
\makecell{\textbf{Outcome}\\\textbf{pathways}} &
\makecell{\textbf{Semantic}\\\textbf{hierarchy}} &
\makecell{\textbf{Event}\\\textbf{traceability}} &
\makecell{\textbf{Multi-scale}\\\textbf{diagnosis}} \\
\midrule

Next-pitch prediction%
~\cite{ganeshapillai2012predicting,hamilton2014pitchprediction,yu2022attentionlstm,lee2022pitchlocation} &
Pitch-choice prediction &
\cmark &
\xmark &
\pmark &
\xmark &
\xmark &
\xmark &
\xmark \\

MDP / game-theoretic sequencing%
~\cite{sidhu2014moneybarl,melville2023gametheory} &
Strategic pitch selection &
\cmark &
\xmark &
\cmark &
\pmark &
\xmark &
\xmark &
\xmark \\

Pitch tunneling / trajectory similarity%
~\cite{long2017pitchtunnels,kagan2017statcast} &
Pairwise pitch execution &
\pmark &
\cmark &
\xmark &
\pmark &
\xmark &
\pmark &
\xmark \\

Pitch-sequence graph / motif studies%
~\cite{prasad2021graphsequencing,park2026pitchmotifs} &
Sequence-structure discovery &
\cmark &
\pmark &
\pmark &
\pmark &
\pmark &
\pmark &
\pmark \\

Pitch-value / outcome models%
~\cite{healey2019pitchvalue} &
Pitch-quality estimation &
\xmark &
\pmark &
\pmark &
\cmark &
\xmark &
\pmark &
\xmark \\

Counterfactual sequence optimization%
~\cite{takamido2026counterfactual} &
Strategy optimization &
\cmark &
\xmark &
\cmark &
\cmark &
\pmark &
\pmark &
\xmark \\

\midrule

\textbf{\system{} (ours)} &
\textbf{Pitching-strategy diagnosis} &
\textbf{\cmark} &
\textbf{\cmark} &
\textbf{\cmark} &
\textbf{\cmark} &
\textbf{\cmark} &
\textbf{\cmark} &
\textbf{\cmark} \\

\bottomrule
\end{tabularx}
\end{table*}

\section{Background and Problem Formulation}
\label{sec:background}
\subsection{Pitching as Sequential Event Data}

A plate appearance (PA) is a variable-length interaction in which a pitcher selects and executes an ordered sequence of pitches against a batter. 
Each pitch is simultaneously a strategic decision, a physical action, and an observed event whose meaning depends on its nominal type, physical execution, preceding pitches, and game context. 
We represent the $i$-th pitch event as $e_i=(s_i,r_i,c_i,o_i)$, where $s_i$ denotes nominal pitch identity, $r_i$ continuous physical execution, $c_i$ information available before the pitch, and $o_i$ post-pitch annotations. 
Physical execution includes release conditions, velocity, movement, reconstructed three-dimensional trajectory, and plate location; context includes count, handedness matchup, base-out state, inning, and score situation; and post-pitch annotations include batter response, contact quality, and run-value change. 
Outcomes describe what followed an execution but do not determine pitch identity: nominally identical pitches may follow different trajectories and produce different responses, while similar outcomes may arise from different physical and sequential mechanisms.

A plate appearance is an ordered sequence $P=(e_1,e_2,\ldots,e_T)$, where $T$ varies across plate appearances. 
Consecutive valid events define observed pitch-to-pitch relations, and these local sequences are nested within games and longer temporal windows. 
Pitch-tracking data are therefore naturally hierarchical spatio-temporal event data rather than independent rows. 
This structure creates a resolution trade-off: coarse states such as pitch type are compact and repeatedly observed but collapse within-type physical variation, whereas adding trajectory, location, context, and longer pitch histories rapidly fragments the state space. 
A useful representation must therefore preserve the underlying physical events while adapting the resolution at which recurring sequence structure is summarized.

\subsection{Related Work}

\paragraph{Pitch prediction and strategic decision modeling.}
Pitch-prediction studies estimate the next pitch type or location from pitcher tendencies, batter information, count, and previous pitches~\cite{ganeshapillai2012predicting,hamilton2014pitchprediction,lee2022pitchlocation,yu2022attentionlstm}. 
Reinforcement-learning, Markov-decision, and game-theoretic approaches further model pitch selection as a sequential decision problem~\cite{sidhu2014moneybarl,melville2023gametheory}. 
These studies establish that pitch choice depends on prior actions and context, but primarily target prediction or strategy optimization rather than retrospective organization of observed execution.

\paragraph{Pitch sequences and graph representations.}
Pitch sequences have been studied through transition structures, recurring motifs, and directed graph representations~\cite{bock2015sequencecomplexity,prasad2021graphsequencing,park2026pitchmotifs}. 
Such models capture pitch order efficiently, but states are commonly defined by nominal pitch labels or other discrete categories, which may merge physically different realizations of the same sequence. 
Increasing state detail or path length can recover specificity, but at the cost of rapidly decreasing statistical support.

\paragraph{Trajectory and outcome analysis}
PITCHf/x and Statcast enable detailed characterization of release position, velocity, movement, location, and pitch flight~\cite{kagan2017statcast,lee2025analyzing}, while pitch-tunneling analyses examine how consecutive pitches converge and diverge through flight~\cite{long2017pitchtunnels}. 
Pitch-level models additionally estimate effectiveness from pitch type, physical execution, location, and game state~\cite{healey2019pitchvalue}. 
These approaches capture complementary aspects of pitching, but trajectory geometry or outcome prediction alone does not expose the recurring ordered structures and temporal contexts through which an execution acquires strategic meaning.

\paragraph{Positioning of \system{}}
As summarized in Table~\ref{tab:related_work_gap}, existing approaches typically capture only a subset of sequence structure, physical execution, context, outcome pathways, and event-level traceability. 
Sequence models emphasize order, trajectory analyses physical execution, and outcome models effectiveness. 
\system{} addresses the representation problem that arises when these elements must be analyzed jointly at scale: it preserves exact physical events and observed sequence relations while supporting semantic and temporal aggregation, event-level traceability, and multi-scale diagnosis without forcing every analysis into a single fixed state space. 
Our focus is therefore on representations that preserve explicit sequence
structure and recoverable supporting events. Predictive sequence encoders
optimize a different objective, and \system{} does not claim superiority as
a predictive architecture.

\subsection{Problem Formulation}

Given pitch-tracking records for pitcher $p$ over analysis period $\tau$, let
$\mathcal{D}_{p,\tau}=\{P_1,P_2,\ldots,P_N\}$ denote the observed plate-appearance sequences. 
Our goal is to construct an attributed hierarchical graph $G_{p,\tau}=(V^{\mathrm{evt}} \cup V^{\mathrm{sem}} \cup V^{\mathrm{time}}, E^{\mathrm{seq}} \cup E^{\mathrm{sem}} \cup E^{\mathrm{time}}),$ where $V^{\mathrm{evt}}$ contains exact pitch events, $V^{\mathrm{sem}}$ semantic index nodes, and $V^{\mathrm{time}}$ temporal-scope nodes. 
The corresponding edge sets encode observed pitch order, semantic membership, and temporal containment.

The canonical object in $G_{p,\tau}$ is the exact pitch event. 
Continuous trajectory and plate location remain attached to each event rather than being replaced by a discrete state; semantic nodes provide alternative resolutions for aggregation and drill-down, while temporal nodes organize the same events across plate appearances, games, rolling windows, and seasons. 
Post-pitch outcomes remain annotations and are excluded from event identity and semantic state construction.

Given $G_{p,\tau}$, the primary diagnostic task is to identify recurring ordered patterns and characterize their execution, contextual use, temporal scope, and observed response pathways while retaining support and lineage to the underlying games, plate appearances, transitions, and pitches. 
Because highly detailed or long paths may occur only a few times, the most specific representation is not always statistically reliable. 
When repeated evidence is insufficient, the representation therefore backs off to a shorter or coarser description rather than elevating a nearly unique execution into a stable strategy pattern.

Accordingly, \system{} is designed around three requirements: \emph{event fidelity}, so that exact physical executions and observed adjacencies remain recoverable; \emph{support-adaptive resolution}, so that sequence length and semantic detail reflect repeated evidence; and \emph{multi-scale traceability}, so that aggregate patterns remain connected to their temporal occurrences and underlying events. 
The objective is not causal identification or a universally superior predictive model, but a shared graph representation for support-aware, multi-scale analysis of observed pitching strategy. 
Prediction and post-execution modeling are used only as auxiliary evaluations under task-specific information constraints. 
Section~\ref{sec:system_design} describes how \system{} instantiates these principles.

\begin{figure*}[t!]
    \centering
    \includegraphics[width=.8\linewidth]{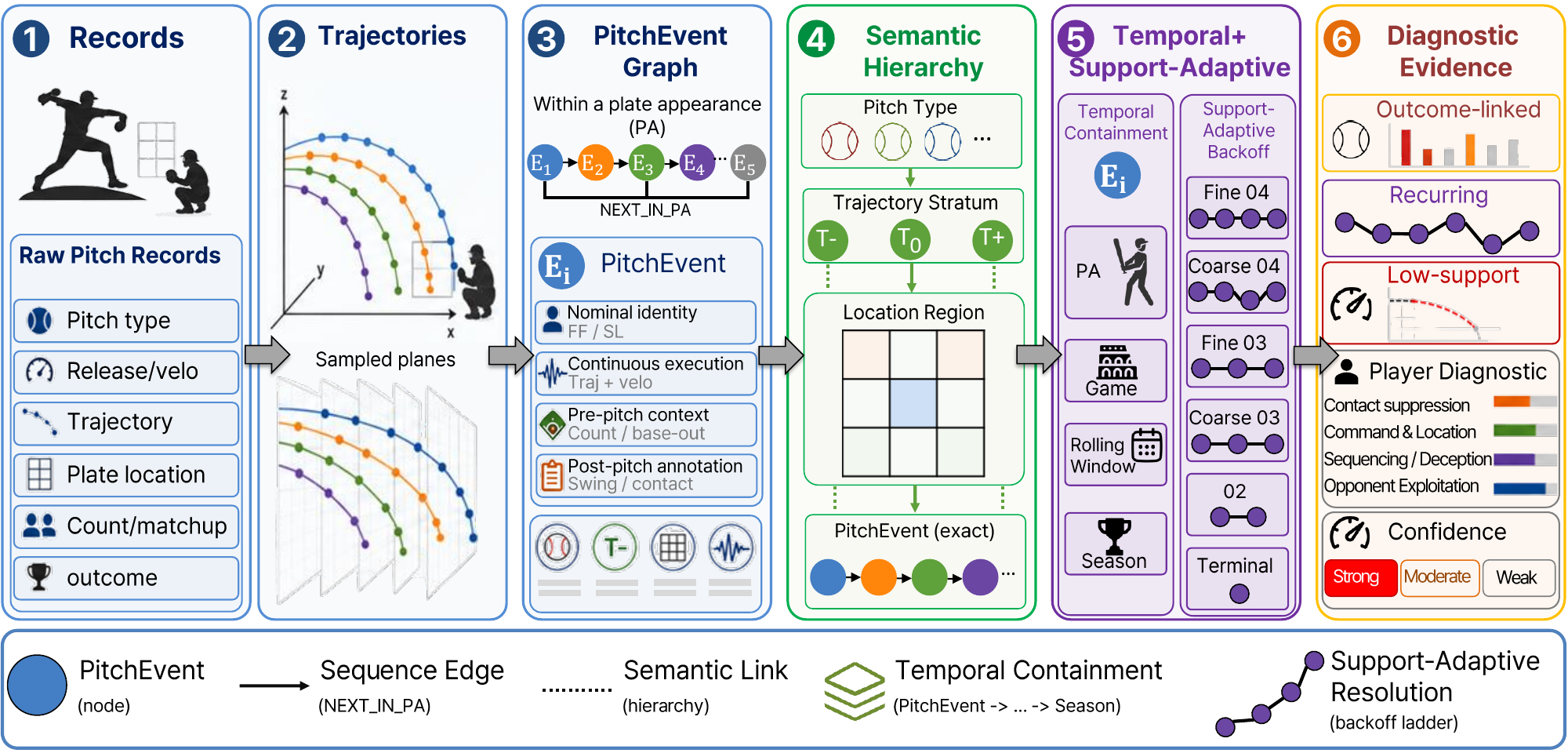}
    \caption{Overview of \system{}. Exact pitch events preserve physical
    execution and within-plate-appearance order. Semantic and temporal
    hierarchies organize the same events at multiple resolutions, while
    support-adaptive paths balance sequence specificity with repeated
    evidence for multi-scale diagnosis.}
    \label{fig:system_overview}
\end{figure*}

\section{System Design}
\label{sec:system_design}

\system{} represents pitching strategy as a hierarchy over exact observed
pitch events rather than as a graph with one fixed discrete state space.
As illustrated in Figure~\ref{fig:system_overview}, each pitch remains an
individual \textsc{PitchEvent} with its continuous execution and pre-pitch
context, while consecutive pitches within a plate appearance (PA) form
directed sequence edges.
Semantic links provide different levels of physical resolution, and temporal
links organize the same events across PAs, games, rolling windows, and seasons.
When detailed multi-pitch patterns lack sufficient repeated support,
\system{} backs off to a shorter or coarser representation without discarding
the underlying physical events.

The key distinction is between \emph{event fidelity} and
\emph{analytical resolution}: the former preserves what physically occurred,
whereas the latter determines how specifically a recurring pattern can be
reported from the available evidence.

\subsection{Exact Pitch Events and Sequence Relations}
\label{sec:pitch_event_graph}

Following Section~\ref{sec:background}, each valid observed pitch $i$
corresponds to one event vertex $v_i$ with
$x_i=[s_i,r_i,c_i]$, where $s_i$ is nominal pitch identity, $r_i$ continuous
physical execution, and $c_i$ pre-pitch context.
The physical and contextual variables follow the definitions in
Section~\ref{sec:background}.
Post-pitch batter response, contact quality, and run-value variables are stored
separately as annotations $o_i$ and do not determine event identity or
semantic membership.

The event is \emph{exact} in the sense that it remains in one-to-one
correspondence with an observed pitch rather than being replaced by a
pitch-type node, trajectory centroid, or clustered state.
For consecutive valid pitches $i-1$ and $i$ within the same PA, we add a
directed sequence edge
\begin{equation}
e_i^{\mathrm{seq}}=(v_{i-1},v_i).
\label{eq:sequence_edge}
\end{equation}
The edge preserves observed pitch order together with pairwise execution
descriptors such as changes in velocity, release position, trajectory,
late-flight separation, and plate location.
Thus, nominally identical pitch-type transitions can remain distinguishable
through their physical realizations.
Invalid intermediate observations break the sequence rather than inducing
an artificial adjacency.

Exact events and their sequence edges form the canonical backbone of
\system{}.
The semantic and temporal structures introduced below index and aggregate
this backbone while retaining links to the original pitches.

\subsection{Continuous Execution and Semantic Hierarchy}
\label{sec:semantic_hierarchy}

For each pitch, \system{} reconstructs its three-dimensional flight from
Statcast kinematic measurements:
\begin{equation}
\mathbf{r}_i(t)
=
\mathbf{r}_{0,i}
+
\mathbf{v}_{0,i}t
+
\frac{1}{2}\mathbf{a}_i t^2 .
\label{eq:pitch_trajectory}
\end{equation}
The reconstructed curve is sampled at fixed locations along the flight path
to obtain a compact trajectory representation.
The sampled trajectory, release conditions, and continuous plate coordinates
remain attached to the exact event throughout the analysis.

Continuous execution is not replaced by a single discrete trajectory state.
Instead, \system{} provides semantic resolutions ranging from pitch type,
through within-type trajectory and location refinements, to the exact event.
These levels form an analytical index over the same pitches rather than a
causal or generative hierarchy.
A coarse view can therefore summarize repeated pitch-type transitions, while
a finer view can reveal the physical executions and locations supporting
those transitions.

Trajectory variation is defined separately within each pitch type.
Let $\widetilde{\mathbf q}_i$ denote the standardized sampled trajectory of
pitch $i$ under a historical reference distribution.
For pitch type $k$, let $\mathbf u_k$ be the first principal direction of the
corresponding within-type trajectory distribution.
We define
\begin{equation}
z_i
=
\mathbf u_k^\top \widetilde{\mathbf q}_i .
\label{eq:trajectory_score}
\end{equation}
Frozen reference tertiles of $z_i$ define three trajectory strata,
$T^{-}$, $T^{0}$, and $T^{+}$.
The standardization, projection, and thresholds are estimated before the
diagnostic window and then held fixed, giving the strata a consistent meaning
across time.
They describe relative within-type trajectory variation rather than pitch
quality or effectiveness.

Plate endpoints are also associated with an interpretable location region,
while their original continuous coordinates are retained.
Game variables such as count, handedness, base-out state, inning, and score
remain conditioning attributes rather than default semantic-state components.
Crossing all physical and contextual variables into a single state would
rapidly fragment recurring sequences; \system{} instead preserves these
variables while allowing the analytical resolution to vary independently.

\subsection{Multi-Scale Strategy Graphs}
\label{sec:temporal_hierarchy}

Pitching structure may be local to a PA, recur within a game, emerge over
recent games, or characterize a season.
\system{} therefore links each exact event to its enclosing PA and game and
organizes games into rolling windows and seasons.
These temporal levels do not create new physical observations; they determine
which events are summarized at a particular analytical scope.

Let $\phi_r(v_i)$ map exact event $v_i$ to its semantic state at resolution
$r$.
For pitcher $p$ and temporal scope $\tau$, the weight of aggregate transition
$(u,v)$ is
\begin{equation}
w_{uv}^{p,\tau,r}
=
\sum_{i\in\mathcal{T}_{p,\tau}}
\mathbb{I}
\left[
\phi_r(v_{i-1})=u,\,
\phi_r(v_i)=v
\right],
\label{eq:aggregate_transition}
\end{equation}
where $\mathcal{T}_{p,\tau}$ contains valid within-PA transitions made by
pitcher $p$ in scope $\tau$.
Changing $r$ changes the physical resolution of the graph, while changing
$\tau$ changes its temporal scope.

Aggregate nodes and edges retain references to their constituent events and
sequence edges.
Their contexts, pairwise execution, and post-pitch annotations can therefore
be summarized without losing event lineage.
A season-level pattern can be localized to supporting games and PAs and then
inspected as individual physical trajectories.
Rolling-window analyses use only completed games available within the
corresponding window.

\subsection{Support-Adaptive Strategy Paths}
\label{sec:support_adaptive_paths}

Pairwise transitions capture immediate pitch order, but recurring pitching
patterns may span longer sequences.
\system{} therefore considers paths of two to four consecutive pitches within
a PA.
For a length-$m$ path ending at pitch $i$ under semantic resolution $r$,
\begin{equation}
\pi_i^{(m,r)}
=
\left(
\phi_r(v_{i-m+1}),
\ldots,
\phi_r(v_i)
\right).
\label{eq:strategy_path}
\end{equation}
For example, a three-pitch path contains pitches $i-2$, $i-1$, and $i$ in
their observed order, and paths never cross PA boundaries.

Longer paths capture more sequential context, while trajectory-refined states
capture more physical specificity.
Both reduce recurrence, creating a trade-off between descriptive detail and
statistical support.
\system{} addresses this trade-off with support-adaptive variable-order
selection.
Each candidate path must satisfy both a minimum number of occurrences and a
minimum number of distinct supporting games.
At each path length, a trajectory-refined representation is considered before
its pitch-type counterpart; if neither is supported, the procedure backs off
to the next shorter suffix.
If no multi-pitch candidate is supported, the target pitch type serves as the
final fallback.

Let $\mathcal{R}$ denote this ordered set of candidate representations.
The selected representation is
\begin{equation}
\rho_i
=
\underset{(m,r)\in\mathcal{R}}{\operatorname{first}}
\left\{
(m,r)
\;\middle|\;
n\!\left(\pi_i^{(m,r)}\right)\ge n_{\min},
\;
g\!\left(\pi_i^{(m,r)}\right)\ge g_{\min}
\right\},
\label{eq:support_adaptive_routing}
\end{equation}
where $n(\cdot)$ is occurrence support and $g(\cdot)$ is the number of
distinct supporting games.

Crucially, backoff changes the resolution of the statistical claim rather
than the information stored in the graph.
A path reported only at pitch-type resolution still retains the trajectories,
locations, contexts, temporal occurrences, and exact sequence edges of all
supporting pitches.
Location and game context can therefore be used for conditioning and
drill-down without being crossed into every default path identity.

\subsection{Diagnostic Evidence and Reporting}
\label{sec:diagnostic_output}

The primary output of \system{} is a retrospective diagnosis of recurring
pitching structure.
We use \emph{motif} to denote a path that satisfies the recurrence criteria
above and is reported as part of a pitcher-level analysis.
A physically distinctive but rarely observed sequence remains inspectable,
but is not treated as evidence of a stable recurring pattern.

For each motif, \system{} records its support across games, temporal
distribution, contextual usage, physical execution, and exact event lineage.
Post-pitch annotations are examined only after the structural motif has been
defined, separating the existence of a recurring pattern from its observed
effectiveness.
Outcome comparisons also respect their relevant populations: whiff evidence
is evaluated among swings, while contact-quality evidence is evaluated among
balls put in play.
When temporal validation is available, the direction of an observed outcome
association is additionally checked outside the discovery observations.

The resulting reports distinguish three levels of evidence.
A recurring motif with a directionally consistent outcome association is
reported as \emph{outcome-linked diagnostic evidence}.
A motif that recurs but has weak or unstable outcome differences remains a
\emph{recurring strategy pattern} without a strong effectiveness claim.
A detailed execution without sufficient recurrence remains an
\emph{event-level example} rather than being promoted to a stable motif.

These reporting levels change the strength of interpretation, not the
underlying representation: every reported pattern remains traceable to its
supporting games, PAs, sequence edges, and exact pitches.
The resulting evidence is observational, and recurrence or held-out
consistency does not establish causal effects or guarantee future persistence.

\section{Evaluation}
\label{sec:evaluation}

\subsection{Evaluation Questions and Protocol}

We evaluate \system{} around four questions that correspond directly to its design goals.
\textbf{RQ1: Representation fidelity.}
Does \system{} preserve physical execution and ordered structure that are lost in discrete sequence representations?
\textbf{RQ2: Adaptive resolution.}
Can support-adaptive routing retain useful sequence detail without fragmenting the representation into unsupported paths?
\textbf{RQ3: Reliability and traceability.}
Are discovered motifs supported by repeated evidence across games, and can aggregate findings be traced back to exact supporting pitches?
\textbf{RQ4: Diagnostic utility.}
Does the resulting representation support meaningful retrospective analyses across changes, players, time, and context?

We use 3.94 million MLB Statcast regular-season pitches from 2021 through July~3, 2026; the partial 2026 season is denoted 2026*.
The primary evaluation population comprises the 300 pitchers with the largest valid 2025 workloads.
Representation dictionaries, trajectory transformations, and support thresholds are estimated on earlier data and frozen before evaluation, and outcome variables never determine graph states.
Unless stated otherwise, uncertainty is estimated by pitcher-cluster bootstrap.
Player-facing trajectories use unmirrored Statcast coordinates in catcher's view.

Our principal discrete comparator is the Sequence Graph Transform (SGT)~\cite{prasad2021graphsequencing}, which summarizes ordered pitch-type or pitch-type--zone symbols within each plate appearance (PA).
Where appropriate, we also compare pitch mix and fixed- versus variable-order path representations.
For execution reconstruction, each supported path is represented by the mean
continuous execution profile estimated from its training occurrences.
We report $R^2$ between these path-level reconstructions and the exact
execution profiles of held-out pitches; higher values indicate that the
representation groups physically similar executions while retaining coverage.
The evaluation proceeds from representation validity to support-adaptive aggregation, then to reliability and traceability, and finally to diagnostic case studies.

\subsection{Representation Fidelity and Ordered Structure}
\label{sec:eval_representation}

This first experiment asks why a trajectory-aware sequence representation is needed at all.
If a discrete pitch-sequence summary already preserves the relevant structure, then a more elaborate event-level graph would be unnecessary.
We therefore test two points: whether physically different executions can collapse into the same discrete sequence cell, and whether meaningful sequence signal appears only at longer ordered contexts.

\begin{figure}[t]
    \centering
    \includegraphics[width=\linewidth]{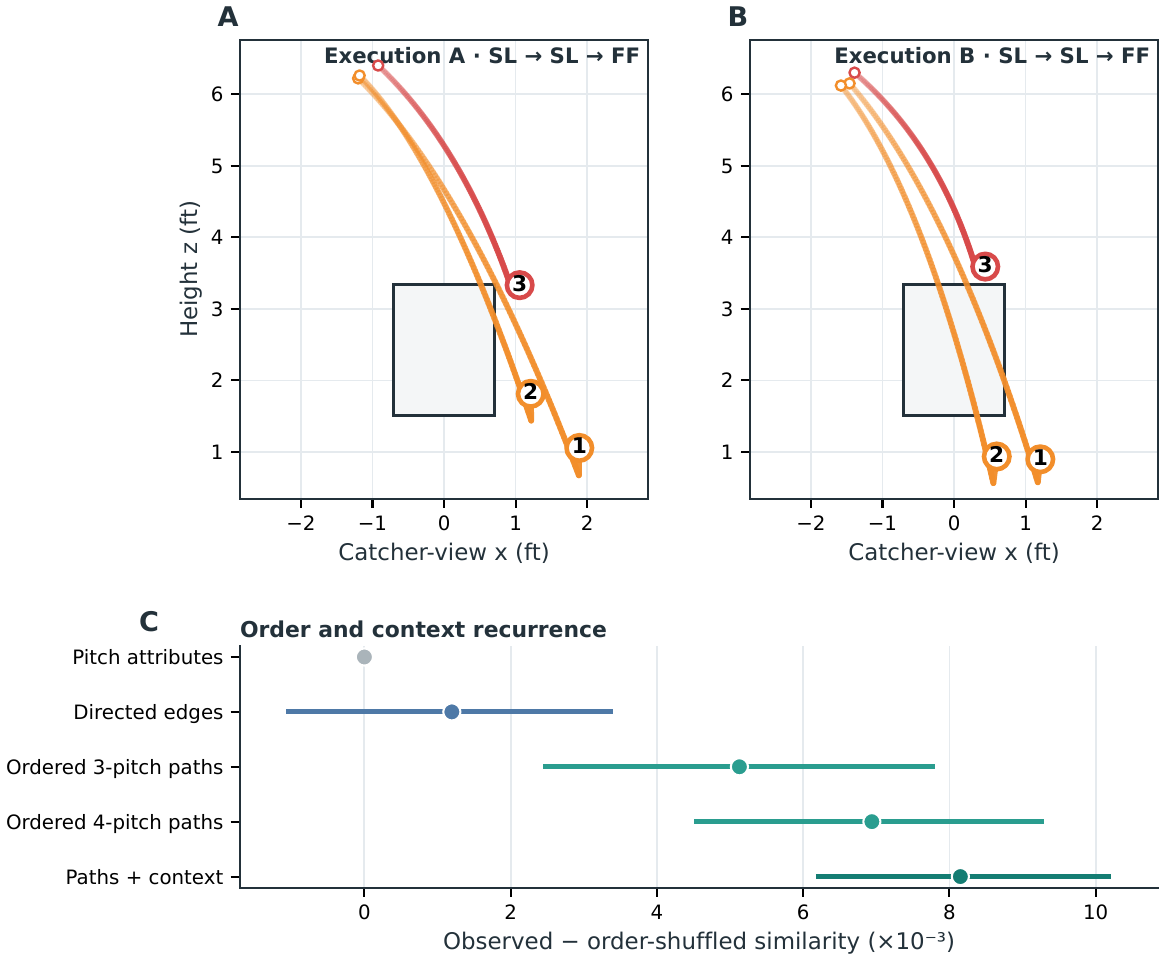}
    \caption{Information retained beyond a discrete sequence.
    (A--B) Two executions assigned the same SGT pitch-type, zone, and count
    cell; numbers indicate pitch order.
    (C) Difference between observed cross-window recurrence and recurrence
    after shuffling pitch order within each PA. Error bars show bootstrap
    confidence intervals.}
    \label{fig:execution-order-context}
\end{figure}

Figure~\ref{fig:execution-order-context}A--B gives a concrete collision example.
The two sequences share the same discrete description
(\texttt{SL--SL--FF} with the same zone/count cell), yet their trajectories are visibly different.
The point of the example is not that every discrete cell is heterogeneous, but that discrete symbols can merge physically distinct executions that a strategy analysis may wish to separate.
This directly motivates the exact-event design of \system{}.

Figure~\ref{fig:execution-order-context}C then asks whether pitch order itself carries information beyond the set of pitches thrown.
We shuffle pitch order within each PA while preserving the observed events.
The result is intuitive and important: shuffling has little effect on isolated event attributes or a single directed transition, but it clearly reduces recurrence for three-pitch, four-pitch, and context-conditioned paths.
Thus, the relevant structure is not simply that one pitch followed another, but that longer ordered subsequences recur in non-random ways.
Together, these results justify a representation that preserves both exact execution and multi-pitch sequence context.

\subsection{Support-Adaptive Resolution}
\label{sec:eval_support}

The next question is whether a highly detailed sequence representation is actually usable at scale.
A graph that preserves fine trajectory detail is only helpful if recurring patterns can still be supported often enough to analyze.
This experiment therefore tests the core design trade-off of \system{}: how much detail can be retained before the representation becomes too sparse.

\begin{figure}[t]
    \centering
    \includegraphics[width=\linewidth]{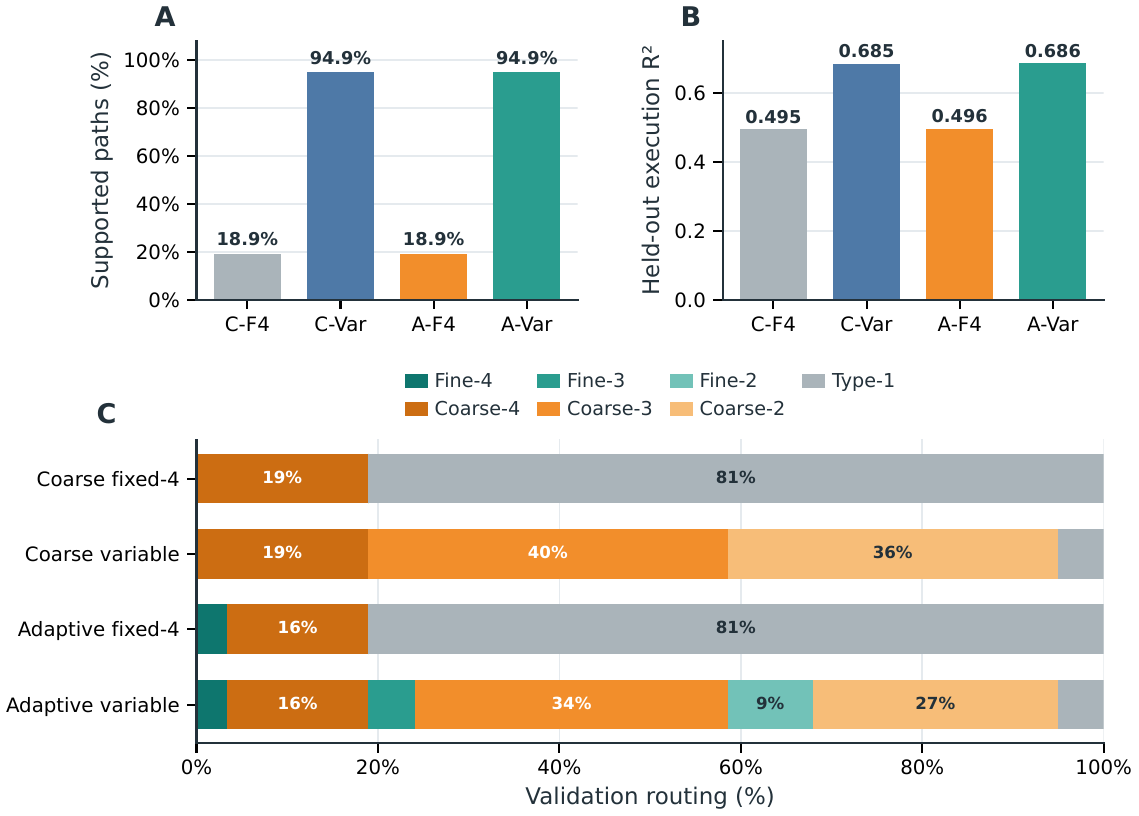}
    \caption{Support and execution fidelity under four hierarchy variants.
    Coarse (C) uses pitch type, adaptive (A) permits qualified trajectory
    substates, F4 fixes every path at order four, and Var backs off through
    shorter suffixes.
    (A) Supported held-out paths.
    (B) Held-out execution reconstruction ($R^2$).
    (C) Selected resolution and order for validation paths.}
    \label{fig:support-adaptive-hierarchy}
\end{figure}

Figure~\ref{fig:support-adaptive-hierarchy} evaluates four alternatives using a chronological split of each pitcher's 2025 games.
A fixed order-four path captures rich local history, but its support collapses: fewer than one in five held-out paths remain supported.
Variable-order backoff resolves this problem, raising coverage to nearly 95\% while also improving held-out reconstruction of physical execution.
Allowing trajectory-qualified substates adds only a small additional gain in reconstruction, showing that the main benefit comes from adapting path length to available support rather than forcing every path into a fine discrete state.

This experiment is central to the paper because it validates the main design decision in Section~\ref{sec:support_adaptive_paths}.
\system{} does not insist that the most detailed path is always the best one.
Instead, it preserves continuous trajectory at the event level and uses a support-qualified variable-order path as the statistical backbone.
Figure~\ref{fig:support-adaptive-hierarchy}C makes this operational: most validation paths are reported at coarse orders two to four, while only a smaller fraction are supported at fine trajectory-refined resolutions.
The result is a representation that is both physically faithful and statistically usable.

\subsection{Reliability, Confidence, and Multi-Scale Traceability}
\label{sec:eval_reliability}

Once a representation can express recurring patterns, the next question is whether those patterns are trustworthy.
A useful diagnostic motif should not be driven by a single game, and a season-level claim should remain traceable to the exact PAs and pitches that support it.
This subsection therefore evaluates both \emph{reliability} and \emph{traceability}.

\begin{figure}[t]
    \centering
    \includegraphics[width=\linewidth]{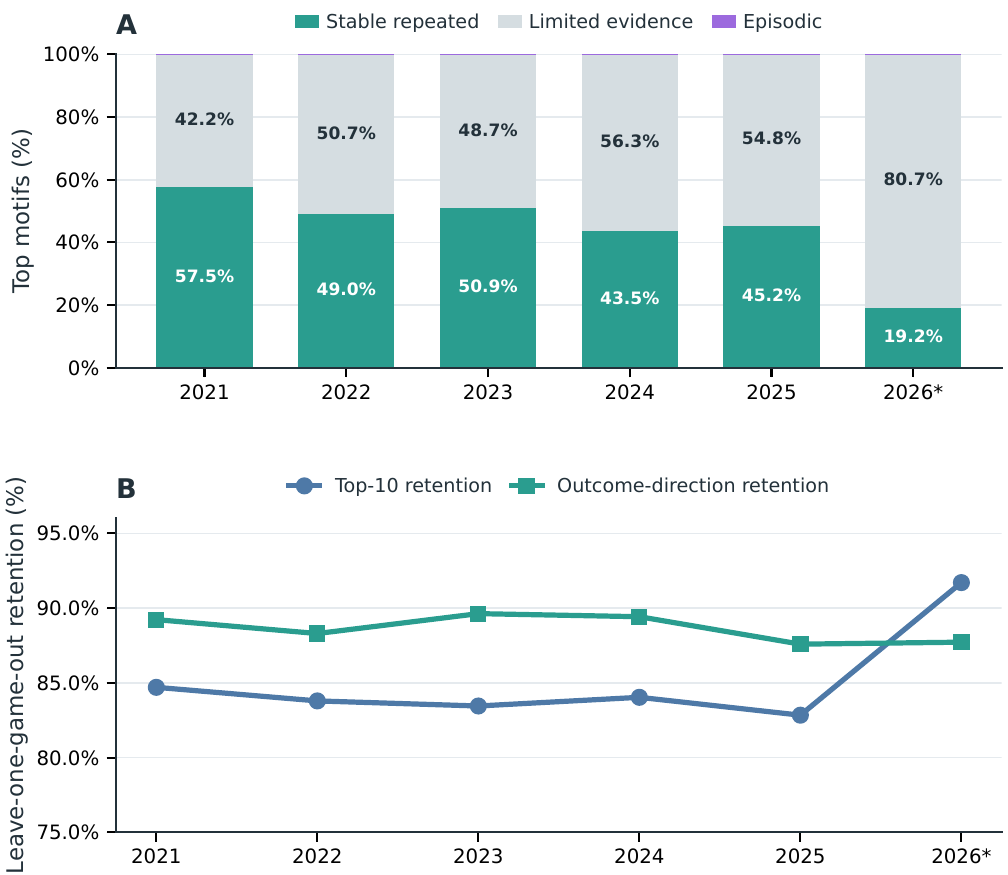}
    \caption{Cross-game reliability of the ten highest-support motifs for each
    pitcher and season.
    (A) Evidence class.
    (B) Retention of top-ten membership and pitcher-beneficial outcome
    direction after removing the motif's highest-support game.
    The 2026* bar reflects a shorter observation window.}
    \label{fig:multiscale-reliability}
\end{figure}

Figure~\ref{fig:multiscale-reliability} asks whether top motifs remain visible after removing their single highest-support game.
They largely do.
Across full seasons, a substantial share of high-support motifs are classified as stable repeated patterns, and most retain both top-ten membership and outcome-direction agreement after the most favorable game is removed.
In partial 2026*, the stable fraction drops sharply and most motifs are labeled limited evidence.
This is the desired behavior.
Rather than over-interpreting short observation windows, \system{} becomes more conservative when recurrence evidence is limited.

\begin{figure}[t]
    \centering
    \includegraphics[width=\linewidth]{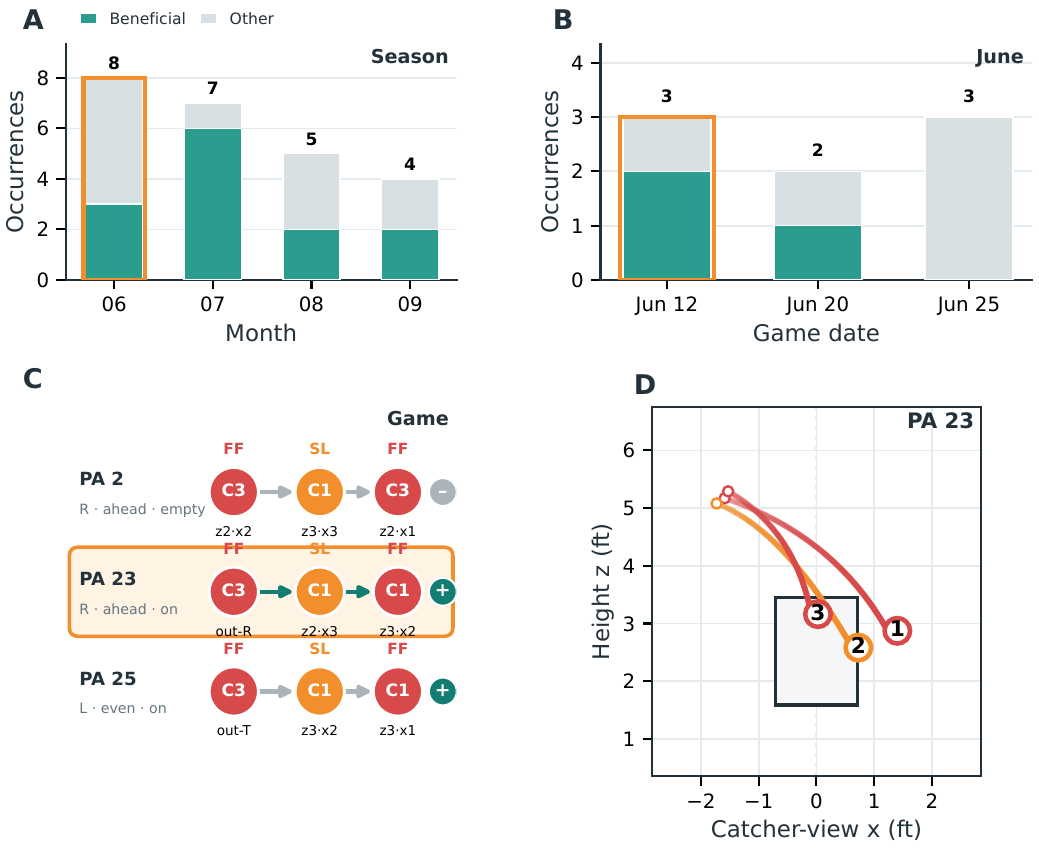}
    \caption{Multi-scale drill-down for Jacob Misiorowski's 2025
    \texttt{FF--SL--FF} motif.
    (A) Monthly support and pitcher-beneficial annotations.
    (B) Support within June games.
    (C) Matching PAs on June~12 with context, trajectory substate, and endpoint.
    (D) Exact catcher's-view trajectories for one occurrence.}
    \label{fig:multiscale-drilldown}
\end{figure}

If Figure~\ref{fig:multiscale-reliability} shows that motifs are not merely artifacts of one game, Figure~\ref{fig:multiscale-drilldown} shows what it means for a motif to remain traceable.
We use Jacob Misiorowski's \texttt{FF--SL--FF} motif because it is frequent enough to support aggregation but still simple enough to visualize clearly.
The figure resolves the same motif from season-level support to monthly counts, then to specific June games, then to the individual PAs on June~12, and finally to the exact catcher's-view trajectories of one occurrence.
This is precisely the intended multi-scale behavior of \system{}: an aggregate strategy pattern remains linked to the physical events from which it was constructed.

The same example also clarifies the meaning of support-adaptive reporting.
The coarse motif is well supported across games, whereas its strict fine realization is not.
Accordingly, \system{} reports the supported coarse pattern but does not discard the trajectory substates, endpoints, or exact pitches.
Backoff therefore weakens the strength of the aggregate claim without deleting the fine-grained evidence.





\begin{table}[t]
\centering
\scriptsize
\caption{Confidence-aware diagnostic examples. $\Delta$ denotes the
outcome-rate difference from the corresponding pitch-type reference.}
\label{tab:diagnostic-confidence}

\setlength{\tabcolsep}{2.5pt}
\renewcommand{\arraystretch}{1.06}

\begin{adjustbox}{width=\columnwidth,center}
\begin{tabular}{llcrrl}
\toprule
Player & Motif &
\makecell{Support\\(occ./games)} &
\makecell{Disc.\\$\Delta$} &
\makecell{Val.\\$n$ / $\Delta$} &
Decision \\
\midrule

S\'anchez
& \texttt{SI--CH--CH}
& 63 / 25
& $-0.8$
& 31 / $-3.7$
& \textsc{Supported} \\

Elder
& \texttt{SI--SL--SL}
& 55 / 22
& $+5.2$
& 7 / $-2.5$
& \textsc{Uncertain} \\

Alcantara
& \texttt{SI--SI--CH}
& 17 / 11
& $-3.0$
& 8 / $-13.7$
& \textsc{Abstain} \\

\bottomrule
\end{tabular}
\end{adjustbox}
\end{table}

Table~\ref{tab:diagnostic-confidence} complements the population-level results with three concrete diagnostic outcomes.
The goal here is not to rank players, but to show how the system reports evidence at different strengths.
Cristopher S\'anchez provides a clear supported case: the motif repeats broadly and retains its direction in validation.
Bryce Elder shows a different situation: the sequence structure repeats, but its outcome association does not remain stable, so the system retains the motif while downgrading the claim.
Sandy Alcantara illustrates abstention: the candidate pattern is inspectable but does not have enough repeated support to justify a stable diagnosis.
This table is important because it shows that \system{} does not force every interesting sequence into a strong claim.

\subsection{Retrospective Change Localization}
\label{sec:eval_change}

A further motivation for the framework is retrospective diagnosis of how a pitcher's style changes.
This requires more than detecting that aggregate statistics moved; it requires localizing what changed and which events support that conclusion.
We therefore evaluate change localization in both a controlled setting and a natural retrospective setting.

\begin{figure}[t]
    \centering
    \includegraphics[width=\linewidth]{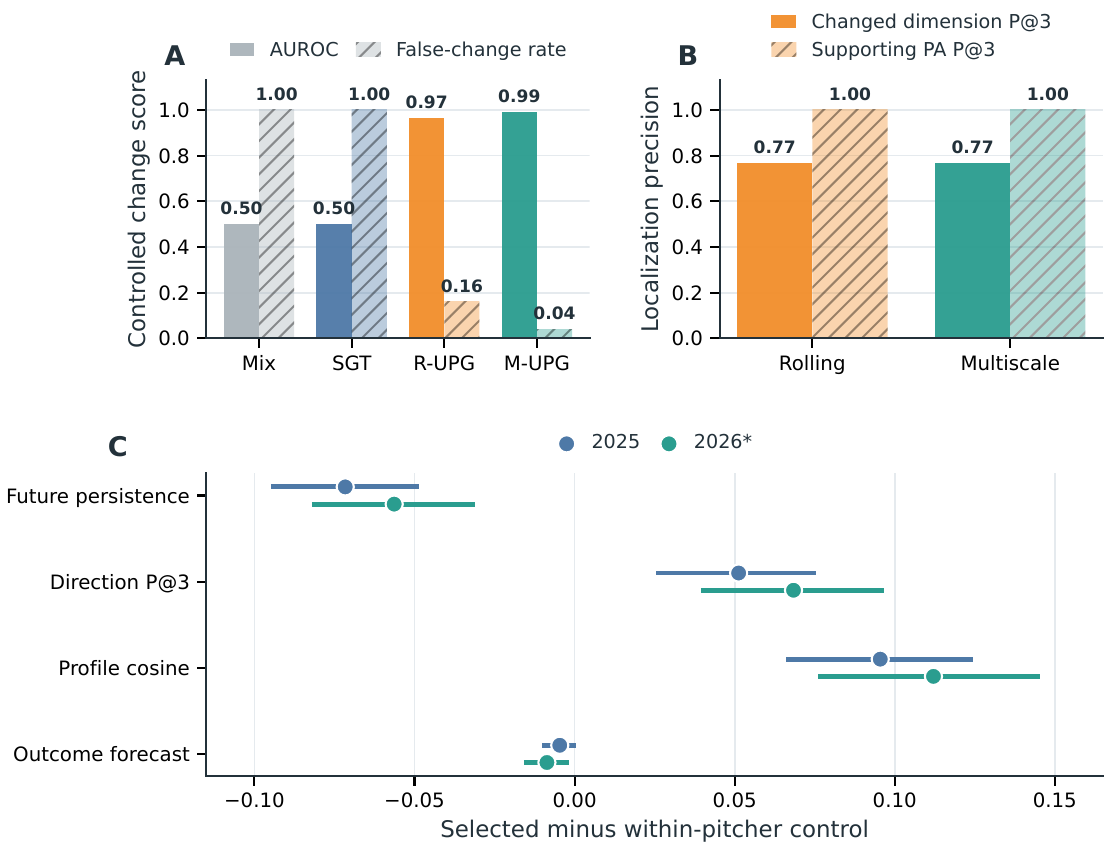}
    \caption{Change detection and diagnostic guardrails.
    (A) Detection of a controlled execution-only change and false-change rate.
    (B) P@3 for the known changed dimensions and supporting PAs.
    (C) Gain over within-pitcher control boundaries for natural changes;
    intervals are pitcher-bootstrap confidence intervals.}
    \label{fig:change-validity}
\end{figure}

In the controlled study, the pitch type, zone, and count signature are held fixed while the continuous execution distribution is changed at a known game boundary.
This design isolates what \system{} is meant to detect: execution-level change that is invisible to coarse summaries.
Figure~\ref{fig:change-validity}A shows that pitch mix and SGT remain at chance, whereas multiscale \system{} achieves near-perfect discrimination.
Figure~\ref{fig:change-validity}B further shows that the method identifies not only that a change occurred, but also the affected dimensions and supporting PAs.
This is the key validity result: the framework can recover localized execution changes when the truth is known.

Figure~\ref{fig:change-validity}C moves to natural retrospective boundaries.
Here the message is intentionally more modest.
Selected boundaries improve reconstruction and directional localization relative to within-pitcher controls, but they do not imply that the change will persist or that future outcomes will improve.
This limitation is important.
The contribution of \system{} is retrospective diagnosis and evidence localization, not a claim of causal discovery or future-performance forecasting.

\subsection{Player-Level, Longitudinal, and Contextual Diagnosis}
\label{sec:eval_player}

The preceding experiments establish that \system{} preserves execution detail, adapts resolution to support, and reports recurring patterns conservatively.
We now show what those properties enable in practice.
The following cases are chosen to illustrate three complementary analytical uses of the framework: comparing different pitchers, tracing one pitcher's reorganization over time, and describing how the same pitcher adapts across matchup contexts.

\begin{figure}[t]
    \centering
    \includegraphics[width=.95\linewidth]{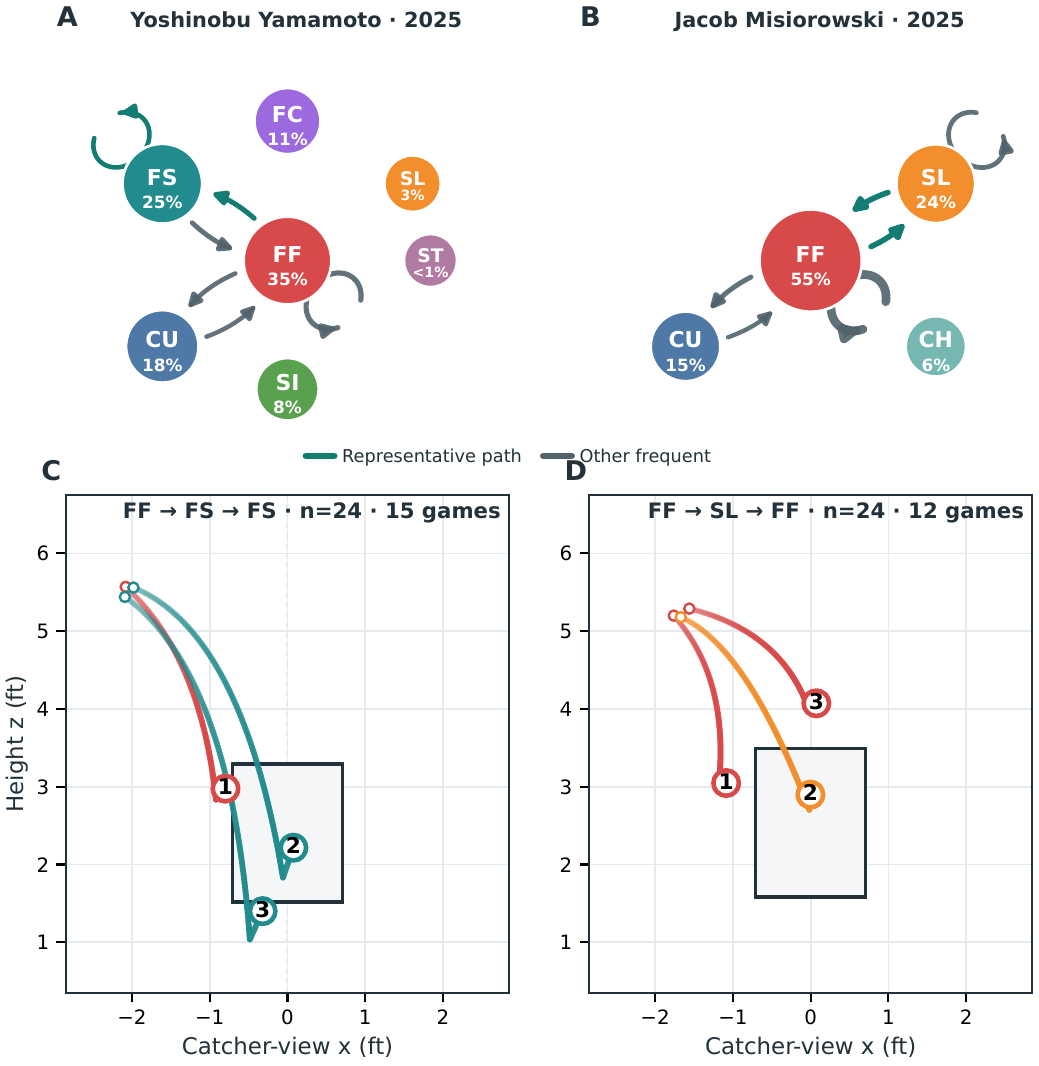}
    \caption{Graph-to-pitch comparison in 2025.
    Node area represents pitch share and edge width uses a common transition
    scale. The lower panels resolve Yoshinobu Yamamoto's
    \texttt{FF--FS--FS} and Jacob Misiorowski's \texttt{FF--SL--FF} motifs
    to representative PAs and exact catcher's-view trajectories.}
    \label{fig:yamamoto-misiorowski}
\end{figure}

\noindent\textbf{Across players.}
We choose Yoshinobu Yamamoto and Jacob Misiorowski because they provide two clearly contrasting organizations of effective pitching.
Yamamoto works from a relatively diverse repertoire with richer mixing among pitch types, whereas Misiorowski builds much of his attack around an unusually concentrated high-velocity fastball--slider backbone.
Figure~\ref{fig:yamamoto-misiorowski} shows that this difference is visible not only in pitch shares and transition structure, but also in the representative executions supporting their motifs.
The case illustrates the intended use of \system{} in player comparison: it characterizes \emph{how} pitchers organize their arsenals, not merely how often they throw each pitch.

\begin{figure}[t]
    \centering
    \includegraphics[width=.95\linewidth]{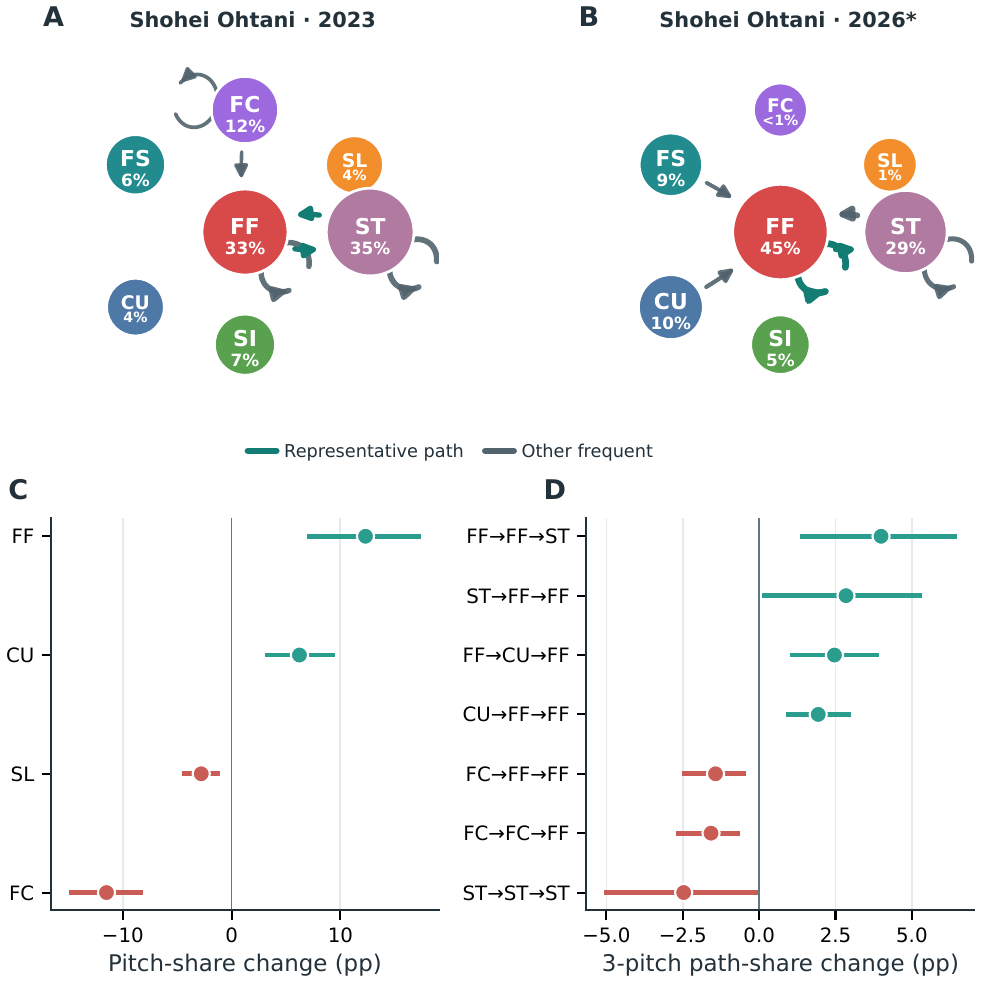}
    \caption{Shohei Ohtani in 2023 and partial 2026*.
    (A--B) Pitch-type graphs.
    (C) Pitch-share changes.
    (D) Largest three-pitch path-share changes.}
    \label{fig:ohtani-longitudinal}
\end{figure}

\noindent\textbf{Across time.}
Ohtani provides a particularly informative longitudinal example because his pitching record contains a clear interruption after 2023 and a later return with a visibly reorganized style.
This makes him an appropriate case for asking not only whether aggregate usage changed, but whether the structure of his sequencing and execution changed as well.
Figure~\ref{fig:ohtani-longitudinal} shows that the difference extends beyond pitch mix:
four-seam usage rises, cutter usage largely disappears, and the dominant three-pitch paths shift toward a different set of recurrent combinations.
The post-return period also shows higher four-seam velocity and improved contact-quality indicators, although not every performance measure improves.
Accordingly, this case is best interpreted as a coordinated reorganization of repertoire, execution, and sequence structure, rather than as a simple claim that his results uniformly improved.

\begin{figure}[t]
    \centering
    \includegraphics[width=\linewidth]{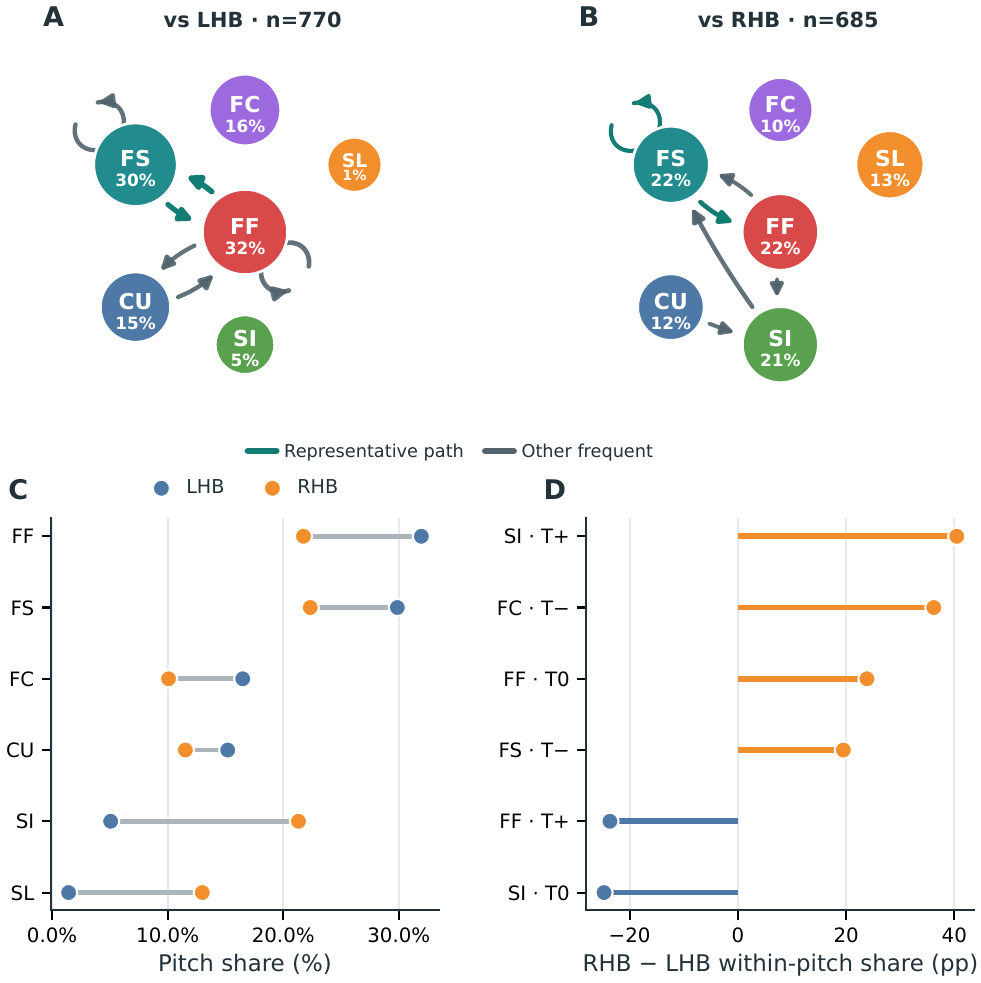}
    \caption{Yamamoto's 2026* strategy by batter side.
    (A--B) Pitch-type graphs against left-handed (LHB) and right-handed
    batters (RHB).
    (C) Pitch-mix comparison.
    (D) Within-pitch trajectory-stratum differences.}
    \label{fig:yamamoto-handedness}
\end{figure}

\noindent\textbf{Across contexts.}
We return to Yamamoto to isolate a different capability of the representation: describing how the same pitcher reorganizes his strategy under different matchup contexts.
Holding the pitcher fixed makes this case complementary to the cross-player comparison above.
Figure~\ref{fig:yamamoto-handedness} shows a clear handedness-conditioned reorganization.
Against left-handed batters, Yamamoto emphasizes four-seam fastballs and splitters; against right-handed batters, sinkers and sliders become much more prominent.
The difference also appears within trajectory strata, indicating that contextual adaptation involves not only which pitches are chosen, but also how they are executed.
This is exactly the kind of structured, context-conditioned diagnosis that a flat pitch-mix summary cannot provide.

\noindent\textbf{Beyond individual cases.}
The preceding figures are intentionally selected to illustrate different analytical questions.
Table~\ref{tab:player-styles} complements them by applying the same hierarchy-native summary to several pitchers with substantially different repertoire breadth and sequence concentration.
Its role is not to declare a best style, but to show that the same representation supports a compact and consistent style description across pitchers.
Table~\ref{tab:player-styles} makes the diversity of graph-native styles explicit.
Misiorowski is highly concentrated around a four-seam backbone and a small set of recurring paths, whereas Messick and Griffin distribute usage across broader repertoires with less concentrated path structure.
Yamamoto and Ohtani occupy intermediate positions with different leading motifs and trajectory-refined backbone executions.
The value of the table is therefore not in any single number, but in showing that \system{} supports a coherent multi-player style vocabulary.

\begin{table}[t]
\centering
\scriptsize
\caption{Hierarchy-native player style signatures. Eff. is effective
repertoire; Top-10 is path concentration; Sw. is mean pitch-type
switches per three-pitch path.}
\label{tab:player-styles}

\setlength{\tabcolsep}{2.2pt}
\renewcommand{\arraystretch}{1.06}

\begin{adjustbox}{width=\columnwidth,center}
\begin{tabular}{lclccr}
\toprule
Player &
Eff. &
Backbone &
\makecell{Leading\\3-pitch path} &
\makecell{Top-10\\/ Sw.} &
Execution \\
\midrule

Yamamoto
& 5.43
& FF 27\%
& \texttt{FS--FF--FS} 3.2\%
& 24.4 / 1.65
& \texttt{FF|T+} 49.6\% \\

Misiorowski
& 3.02
& FF 63\%
& \texttt{FF--FF--FF} 28.8\%
& 67.2 / 0.96
& \texttt{FF|T0} 51.7\% \\

Ohtani
& 4.03
& FF 45\%
& \texttt{FF--FF--ST} 8.3\%
& 52.1 / 1.36
& \texttt{FF|T+} 47.3\% \\

Messick
& 5.15
& FF 33\%
& \texttt{FF--FF--FF} 5.0\%
& 28.8 / 1.49
& \texttt{FF|T-} 71.6\% \\

Griffin
& 6.09
& FC 32\%
& \texttt{FF--FC--FC} 2.4\%
& 17.7 / 1.65
& \texttt{FC|T+} 70.2\% \\

\bottomrule
\end{tabular}
\end{adjustbox}
\end{table}







Taken together, the player-level analyses demonstrate four complementary uses of the same representation:
structural comparison across pitchers, longitudinal reorganization within a pitcher, context-conditioned adaptation, and compact style characterization across multiple players.
Across all of these uses, the key property is unchanged: aggregate graph patterns remain connected to the exact physical events that produced them.

Overall, the evaluation supports a qualified but clear conclusion.
\system{} is not claimed to be universally superior on every compressed-sequence or predictive benchmark.
Its contribution is that continuous execution, ordered structure, temporal scope, support-aware aggregation, diagnostic confidence, and exact event lineage coexist in a single representation.
This makes it possible to move from population-scale summaries to game-level and pitch-level evidence without conflating limited support with strong conclusions.
\section{Discussion}
\label{sec:discussion}
\noindent{\textbf{Observational diagnosis rather than causal attribution.}}
\system{} is built from retrospective observational data and therefore identifies associations rather than causal effects. Pitch outcomes may also depend on factors that are not fully observed in pitch-tracking data, including batter anticipation, pitcher fatigue and condition, umpire decisions, catcher coordination, and game-specific plans. Accordingly, a trajectory motif associated with whiffs or favorable contact should not be interpreted as proving that the motif caused the outcome. Nevertheless, \system{} advances diagnostic resolution by linking trajectory variants, pitch order, game context, and outcome pathways that are collapsed in pitch-mix or aggregate pitch-level summaries.

\noindent{\textbf{From observed trajectories to pitching mechanics.}}
\system{} characterizes how a pitch travels and how it is deployed, but it does not fully explain the biomechanical process that produced that trajectory. Release height, arm slot, deception, spin rate, spin axis, spin efficiency, and pitcher-specific physical constraints may all determine which pitch shapes and sequences are feasible. Future work could integrate biomechanical or pose-tracking measurements with the proposed graph representation. The current analysis should therefore be viewed as generating evidence about observable execution and strategy, rather than directly prescribing mechanical changes.

\noindent{\textbf{Toward actionable and externally validated diagnosis.}}
The current evaluation measures representation quality and diagnostic specificity, but does not directly establish whether the resulting reports improve coaching or player development decisions. Future studies could evaluate the reports with pitchers, coaches, and analysts, and prospectively examine whether strategy or training changes based on identified motifs produce the expected effects. Broader league-wide, cross-season, and cross-league evaluations would also clarify how well the framework transfers across competition levels and tracking systems.
\section{Conclusion}
\label{sec:conclusion}

We presented \system{}, a hierarchical graph framework for retrospective analysis of pitching strategy.
\system{} preserves exact pitch events and continuous execution, organizes them across semantic and temporal scales, and uses support-adaptive paths to balance sequence specificity with repeated evidence.
Using 3.94 million MLB Statcast pitches, we showed that discrete sequence representations can miss meaningful execution and ordered structure, while \system{} supports reliable multi-scale drill-down, retrospective change localization, and player-level diagnosis with exact event traceability.
More broadly, these principles may be useful for other sequential-event domains that combine continuous observations with recurring discrete structure.
The framework is intended for observational diagnosis rather than causal inference or future-performance prediction.

\bibliographystyle{ACM-Reference-Format}
\bibliography{reference}


\end{document}